\documentclass[a4paper,twoside]{article}
\newcommand{\affil}[1]{$^{\rm #1}$}
\usepackage[authoryear]{natbib}
\bibpunct{(}{)}{;}{a}{}{,}
\usepackage{graphicx}
\date{} 
\title{\large\bf\flushleft Modelling of laboratory data of bi-directional reflectance of regolith surface containing Alumina}
\author{\parbox{\textwidth}{\flushleft
\vspace{-0.5cm}
{\it C. Bhattacharjee\affil{A,B,C}, D. Deb\affil{A}, H. S. Das\affil{A},  A. K. Sen\affil{A}, and R. Gupta\affil{D}}\\
\vspace{0.4cm}
{\small \affil{A}\,Department of Physics, Assam University, Silchar
              788011, India}\\
 {\small \affil{B}\,Department of Physics, Kokrajhar Govt. College, Kokrajhar
              783370, India}\\             
{\small \affil{C}\, bhattchinmoy@gmail.com}\\
{\small \affil{D}\, IUCAA, Pune, India}\\
}}
\begin{document}
\twocolumn[
\begin{changemargin}{.8cm}{.5cm}
\begin{minipage}{.9\textwidth}
\vspace{-1cm}
\maketitle
%
%
\small{\bf Abstract:} Bidirectional reflectance of a surface is defined as the ratio of
the scattered radiation at the detector to the incident irradiance as a function of geometry.
The accurate knowledge of the bidirectional reflection function (BRF) of layers composed of discrete,
randomly positioned scattering particles is very essential for many remote sensing, engineering,
biophysical applications and in different areas of Astrophysics. The computations of BRF's for plane
parallel particulate layers are usually reduced to solve the radiative transfer equation (RTE) by the
existing techniques. In this work we present our laboratory data on bidirectional
reflectance versus phase angle for two sample  sizes of 0.3 and 1 $\mu m$  of Alumina for  the He-Ne laser at 632.8 nm (red)
and 543.5nm(green) wavelength. The nature of the phase curves of the asteroids depends on the parameters like- particle
size, composition, porosity, roughness etc. In our present work we analyse the data which are being generated using single
scattering phase function i.e. Mie theory considering particles to be compact sphere. The well known Hapke formula will be
considered along with different particle phase function such as Mie and Henyey Greenstein etc to  model the laboratory data
obtained at the asteroid laboratory of Assam University.


\medskip
\medskip
\end{minipage}
\end{changemargin}
]
\small
\section{Introduction}
The study of the light scattering properties of powdered materials is known to be an important tool for
characterizing the physical and compositional properties of asteroids. It is well conceived that asteroids
are covered with finely grained materials known as regolith layers (Hapke 2005). Hence it is imperative that laboratory
based experiments on the asteroid analogues can be compared with the in situ data as well as the theoretical models
can also be tested.  As the phase angle approaches to zero, the brightness of asteroids increases very rapidly, the
phenomenon is termed as opposition effect. The various physical parameters like particle size, porosity, surface
roughness ,thickness of the layer etc are very important and being studied in laboratory by many authors such as
Kamei et al. (1999), Kaasalainen (2003) and Nelson et al. (2000). A large number of literature is available on the physical interpretation of opposition effect based on shadowing and coherent backscattering(Hapke 2002, Shkuratov et al. 2002). But it is difficult to explain with the theoretical models how the opposition
effect depend on physical parameters.

 At large phase angles all the physical parameters cannot be studied efficiently. In spite of
 that certain very important properties like composition, grain size, grain shape etc can be
 studied. The most widely used formula for describing the scattering of light from a particulate
 surface is the Hapke formulae (Hapke 2005) and Lumme \& Bowel formula(1981). It requires at least three unknown parameters, amongst them two become  irrelevant for large phase angles. Recently Hapke et al 2009, compare the ability of  several  radiative transfer models to describe the scattering behavior measured over a wide range of phase angles. Shepard and Helfenstein (2007) studied bidirectional reflectance function for 14 different samples including  4 Al$_2$O$_3$ samples over  a phase angle varied from  3$^0$ to 130$^0$.
Piatek et al. (2004),  measured the variation of reflectance as the phase angle varied from 0.05$^0$ to 140$^0$ for particle  size  ranges from smaller to  larger than the wavelengths.

 In a preliminary work with alumina sample (Deb 2010) for zero tilt and observation wavelength of 632.8 nm, the phase curve was satisfactorily fitted with Hapke formula and Mie theory by varying absorption coefficient \emph{k}. Here, we have included more experimental data at two different particle size of samples (0.3 micron and 1 micron) and observation wavelengths (632.8 nm and 543.2 nm) for different tilt angles to study the theoretical behavior in more detail. In the present work the photometric data at large phase angles for the plane surface of powdered alumina (Al$_2$O$_3$) with 0.3 micron  and 1 micron average particle diameter at the above wavelengths have been generated.
 In the present analysis we have considered Mie theory (single particle scattering) i.e. the particles are compact and
 spherical in shape. We use Mie theory with Hapke formula in Henyey-Greenstein phase function to theoretically calculate
 bidirectional reflectance and model it with the laboratory data thus obtained.

\section{Instrumentation and sample }
 The experiment was carried out with the help of a goniometric device  at    the   department of Physics, Assam
 University, Silchar, India. It consists of two metal arms having a common horizontal axis of rotation. The sample surface
 is placed at the axis of rotation of the arms with the help of three translation stages. A  miniature goniometer acts as a
 tilting device to the sample. The two arms can be rotated by  $\pm 90^0$ from the zenith direction and a tilt up to $\pm 20^0$  can be given
 to the sample from the horizontal position perpendicular to the plane of scattering. We have used  He-Ne laser of  red and green wavelengths  as the source of light and the CCD camera as the detector. The  sample is placed at the common intersection of the axis
 of rotation and axes of the source and detector. A diffuser was placed in front of the CCD to reduce the laser speckles produced by the coherent laser beam on scattering from the rough surface.  It is evident that the diffuser incorporates some uncertainty in emergent angle, to address this criteria  we calculated the solid angle as well as the uncertainty in the emergent angle  as 0.028 Steradian and  $\pm 0.32$ degree respectively.  The intensity at any point on an illuminated area, along a given direction, is defined as the power radiated per unit projected area of illumination, to the direction under consideration, per unit solid angle. In this case, the solid angle feature can be neglected since it is a constant for a particular instrument and finally gets canceled out while taking the ratio.

 The sample used in the present work  is of powdered alumina (Al$_2$O$_3$). We have used two sample of different average value of diameter, 0.3$\mu m$ and 1$\mu m$. Here after
 we shall call 0.3 $\mu m$ as sample-I and 1$\mu m$ as sample-II. At the initial stage the surface roughness is quite high. For the preparation
 of the smooth surface, the sample surface was pressed by a smooth metal plate so that the sample surface takes its smoothness.

\section{Data collection and reduction}
The tilt angle of the sample was set fixed at 0$^0$ first, then varied from $\pm 2^0$ to $\pm 20^0$ at every 2$^0$ interval. For simplicity of the theoretical models, we took tilt angles 0$^0$,10$^0$,20$^0$. The detector angles (e) were kept fixed  at 45$^0$ and 63$^0$  (the sign is positive accounts for forward scattering ref Fig.1) from the Zenith. The angle of incidence ($i$) was varied from 0$^0$ to 63$^0$ in steps of 9$^0$. Hence the phase angles also varied from 45$^0$ to126$^0$. The detector readings were collected at every new angle of incidence, and the images of the sample surface were recorded in the form of FITS image. As the field of view of the detector was larger than the laser spot, geometrical correction (cos i/cos e) was necessary to calculate the intensity values from the detector counts. Corrections for the background were also done for
each observation. The reflectance values were calibrated by using BaSO$_4$ (a standard Lambert surface) at incidence angle 0$^0$ and detector
angle 45$^0$.

\section{Theory}
The bi-directional reflectance r(i,e,g) is defined as the ratio of the reflected intensity (I) to the incidence
irradiance (J) measured for alumina sample is shown in Fig-1 (which shows the experimental set up).

\begin{figure}
\includegraphics[width=84mm]{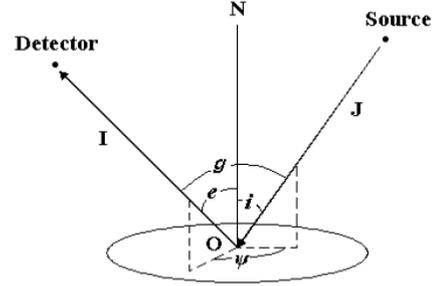}
\caption{Schematic diagram of bidirectional reflectance.}
\end{figure}


 When a beam of collimated light is incident on a rough surface, the Fresnel laws of reflection are not
 obeyed by the reflected light as it gets scattered along all directions throughout the upper hemisphere. The condition $g = i + e$, holds,if the planes of emergence and incidence coincides ($\psi$ = 0 or 180$^0$) and the tilt angle becomes 0$^0$. In the present study  for other tilt angles viz, 10$^0$, 20$^0$ , the phase angle $g \neq i + e$. The intensity of the scattered beam depends on these three angular parameters. The bidirectional reflectance `r' as a  function of i, e and g is given by
 \begin{equation}
 r(i,e,g) = I(i,e,g) / J
\end{equation}

The interrelation among the angle of incidence $i$, detector angle $e$, the phase angle $g$ and the tilt angle $\phi$ is given by,
       \begin{equation}
 cos g = cos i.cos e + sin i . sin e. cos \phi
\end{equation}

\subsection{Mie Theory}
 Mie theory is a single particle light scattering theory, which  was theoretically derived for the solution
 of light scattered from smooth and homogeneous sphere of any size (van de Hulst 1957) . It depends on the complex refractive
 index (n,k) and the size parameter X=2$ \pi a/\lambda$ ,where $a$, $\lambda$ are radius and wavelength of the light respectively.  It is true  fact that Mie theory is applicable only for a single and isolated spherical particle and not directly applicable when there are a number of particles in contact with each other, because, in that case multiple scattering between one particle to another comes across which makes the scattering behaviour complicated. But, the approach considered in this work demands for a 'single particle phase function' into Hapke formula. The calculation of multiple scattering is done by Hapke formula independently. Also, the 'single particle phase function' of a isolated particle and a particle in regolith differes only by a little amount (e.g.Fig.1 of Hapke et al. 2009) which has been neglected in this study.
 To model the laboratory data of bidirectional reflectance, we use Mie theory to calculate single particle albedo  $\omega$
 and the asymmetry parameter $\xi$. It is hardly accepted that the particles of alumina are smooth and homogeneous spheres.
 But Pollack and Cuzzi (1980) suggested that the Mie theory may be used to calculate the scattering properties of  equant
 irregular particles also, provided size parameter X$\le 5$. In the present work the size parameters are 1.49 and 1.73 for sample-1 with red and green wavelengths which also justifies the  above fact.

\subsection{Hapke  Model}
This model describes the scattering of light from a particulate surface, which has been derived from the theory of  r
adiative transfer. The  Hapke  formula mainly has three parameters, i.e. single particle scattering albedo $\omega$,
single particle phase function p(g),opposition surge amplitude B$_0$, opposition surge width h. But for larger
phase angles $>$ 45$^0$, effect of B(g) can be neglected. The Hapke formula is given by,(Hapke 2002, 2005)

\begin{equation}
               r(i,e,g)=(\omega/4\pi){\frac {\mu _0}{\mu _0 + \mu}}[\{1+B(g)\}p(g)+H(\mu _0)H(\mu) -1]
\end{equation}

             where,  $\mu_0$= cos $i$  and $\mu$ =cos $e$  and  $P(g) = (1-\xi ^2 )/(1+2\xi cos g+ \xi ^2)^{3/2}$,
             and $H(\mu _0)=(1+2 \mu _0) / (1+2 \gamma x$), $H(\mu)=(1+2 \mu) / (1+2 \gamma x$) and $\gamma = (1-\omega)^{1/2}$.

It is evident that for brighter surface , the average photon is scattered higher number of times before emerging from the surface  causing the directional effects to be averaged out and multiply scattered intensity distribution to closely approach the isotropic case. The exact  numerical solution for high albedo surface was obtained by Chandrasekhar (1960). The comparison of exact and approximate solution for isotropic scattering has been shown by [ Hapke-1 (1981), in Fig-3,4,5]. In  the same paper Hapke compared the H-functions  versus $\mu$ for several values of single scattering albedo  for Chandrasekhar's exact solution with his approximation(Fig-2)  and found that  the two solutions agree to better than 3\% every where. In actual practice it is seen that single scattering albedo $\omega$ = 1, has never been achieved and a slight decrease in $\omega$  value significantly increases the agreement between exact and approximate solution. It is reported that by Hapke-1(1981) that when $\omega$ = 0.975, the error is only 0.7\% . Therefore it is quite justified that we took Hapke formula for our present analysis. As the theory demands   an arbitrary single particle phase function P(g), we consider an empirical phase function i.e.  Henyey-Greenstein phase function with one term. This  introduces a new unknown parameter $\xi$, known  as  asymmetry parameter. It takes the value between -1 and +1.  These asymmetry factor and single particle scattering albedo  are being calculated from running Fortran Code on Mie theory, Published by Mishchenko et al. (1999), (available online at http://www.giss.nasa.gov/~crmim).

Therefore with the help of  Mie theory in Hapke formula with Henyey-Greenstein phase function (Henyey \& Greenstein 1941), we calculate approximate theoretical bidirectional reflectance of powdered alumina sample. In the next section in results and discussion, we show the nature of
graphs obtained theoretically and compare this with experimentally obtained graphs.

\begin{figure}
\includegraphics[width=75mm]{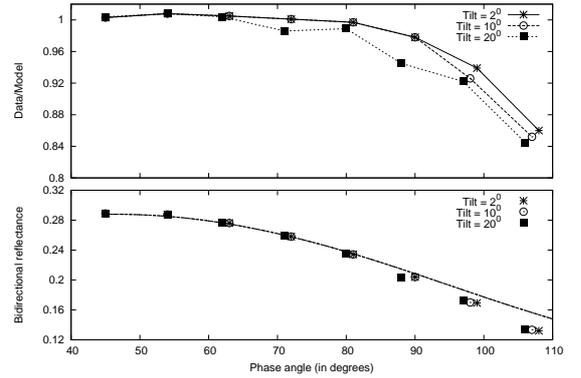}
\caption{ The upper panel shows the matching of Data:Model values ratio to 1.
 The lower panel  gives the bidirectional reflectance vs phase angle  for different tilt angles for    sample-I  at wavelength $\lambda$ = 632.8 nm(e=45$^\circ$). The solid line in the lower panel represents the model. }
\end{figure}

\begin{figure}
\includegraphics[width=75mm]{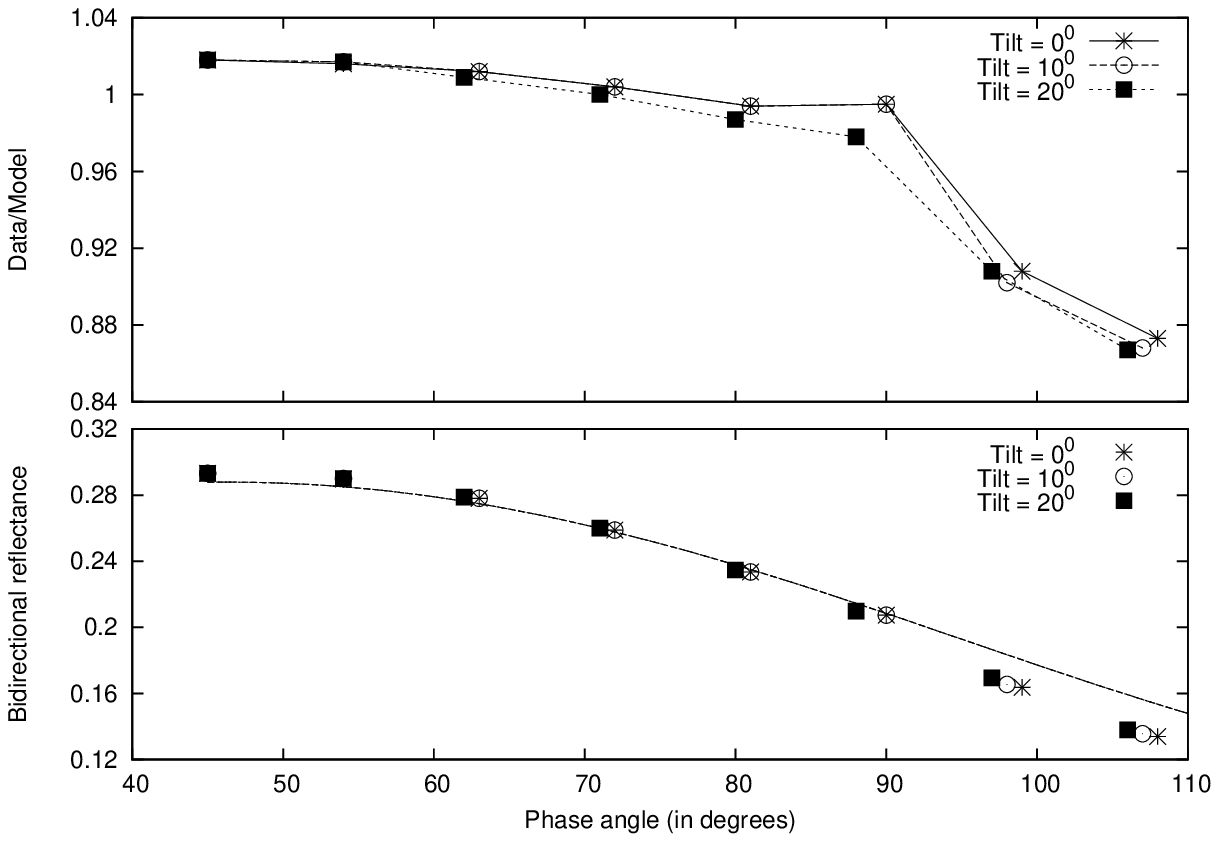}
\caption{ The upper panel shows the matching of Data:Model values ratio to 1.
 The lower panel  gives the bidirectional reflectance vs phase angle  for different tilt angles for    sample-II  at wavelength $\lambda$ = 632.8 nm(e=45$^\circ$).The solid line in the lower panel represents the model. }
\end{figure}

\begin{figure}
\includegraphics[width=75mm]{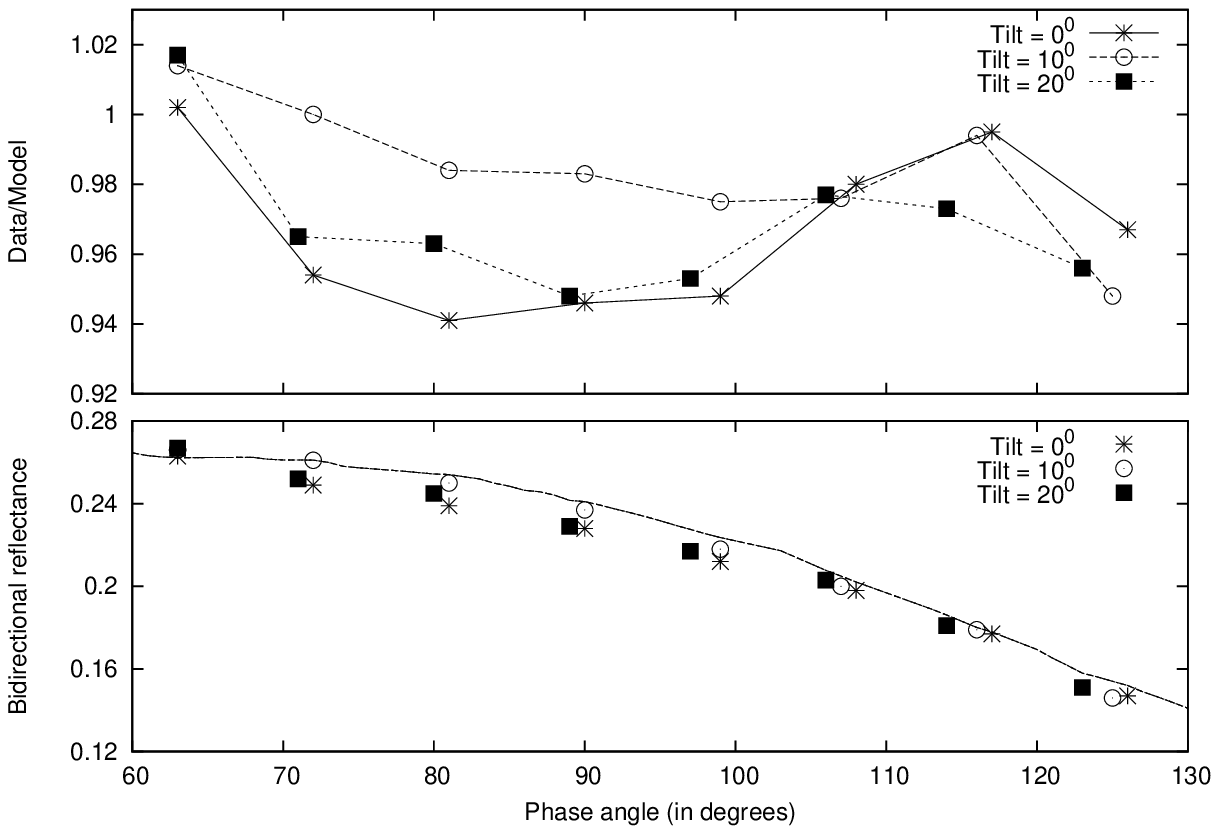}
\caption{ The upper panel shows the matching of Data:Model values ratio to 1.
 The lower panel  gives the bidirectional reflectance vs phase angle  for different tilt angles for    sample-I  at wavelength $\lambda$ = 543.5 nm(e=63$^\circ$). The solid line in the lower panel represents the model. }
\end{figure}

\begin{figure}
\includegraphics[width=75mm]{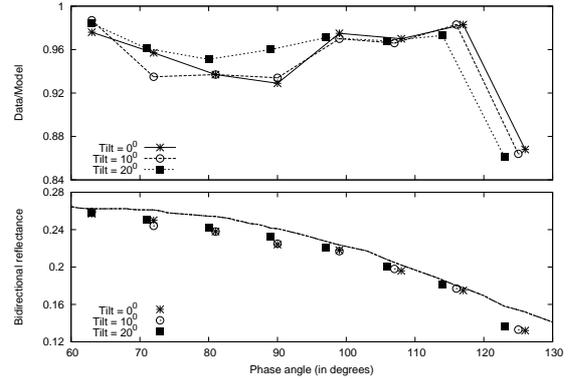}
\caption{ The upper panel shows the matching of Data:Model values ratio to 1.
 The lower panel  gives the bidirectional reflectance vs phase angle  for different tilt angles for    sample-II  at wavelength $\lambda$ = 543.5 nm(e=63$^\circ$). The solid line in the lower panel represents the model. }
\end{figure}



\section{Results and Discussion}
 The refractive index of alumina at 632.8nm is n = 1.766 (Gervais 1991) and absorption coefficient k is known to be very small. For the present fit our free parameter is k, we tried with different values of k and finally we found for our model the appropriate value of k=0.00001. Similarly  for green laser of wavelength  543.5 nm, having n=1.771, the best fit value of k found to be 0.000001 which is comparable with earlier work, reported for tilt angle=0$^\circ$ by Piatek et al(2004).

 Piatek et al.(2004) studied the absolute reflectance versus phase angle for alumina at different phase angles with average particle diameter $\leq$ wavelength, the data thus reported is comparable with the present work for sample-I. In this work we have clearly showed  how the Hapke model can be used to empirically fit the laboratory data not only for zero tilt angle but also for higher tilt angles (eg, 10$^\circ$, 20$^\circ$ etc). Having said that there is a basic difference in our calculation of bidirectional reflectance with that of Piatek et al in their work they kept the angle of incidence fixed  and varied angle of emergence while in this work we have two sets of fixed emergent angles 45$^\circ$, 63$^\circ$ and angle of incidence varied from 0$^\circ$ to 63$^\circ$.

 For sample-II, the average particle size is greater than the wavelength of the laser source. In such condition we have found our  asymmetry parameter $\xi$ is positive which suggests that phase function is forward scattering,for phase angle ranges from 45$^\circ$ to 126$^\circ$. This result is in accordance with other previously reported work which says for non opaque material in a powder the single scattering phase function is forward scattering.(Mishchenko 1994; Mishchenko \& Macke 1997).

 At present we are unable to show how our results is in accordance or in conflict with Shepard and  Helfenstein (2007) , as they tested the significance of Hapke photometric model,  due to non availability of sufficient  photometric data of Al$_2$O$_3$ at average particle diameter grater than wavelength of laser source.

 It is quite obvious that fit to laboratory data of bidirectional reflectance by Hapke model would be better if we can relate the results with physical properties like  porosity, roughness etc. The very fact that we have taken particles to be smooth spheres may incorporate certain uncertainties in modelling as the shape of the particles may be non spherical also.

\section*{Acknowledgments}
We would like to thank the anonymous referee for his/her valuable comments. We  acknowledge  ISRO through its RESPOND programme for financial assistance to carry out this work.


\begin{thebibliography}{}
\bibitem[\protect\citeauthoryear{Chandrasekhar}{2010}]{b1} Chandrasekhar, S., 1960. in \textit{Radiative Transfer}, Dover, New York.

\bibitem[\protect\citeauthoryear{Deb et al.}{2011}]{b1} Deb D., Sen A. K., Das H. S.  and Gupta R., 2011. Adv. Space Res, (in press).

 \bibitem[\protect\citeauthoryear{Gervais et al.}{1986}]{b1} Gervais F, 1991. Handbook of Optical constants of Solid11, Academic press.

 \bibitem[\protect\citeauthoryear{Hapke  et al.}{1981}]{b1} Hapke,  B., 1981(1), JGR, 86,B4, 3039-3054.

 \bibitem[\protect\citeauthoryear{Hapke  et al.}{1986}]{b1} Hapke B., 2002. Icarus, 157:523-34.

 \bibitem[\protect\citeauthoryear{ Hapke  et al.}{1986}]{b1} Hapke B., 2005.  Theory of Reflectance and  Emittance
Spectroscopy, Cambridge University press.
Hapke B..  Theory of Reflectance and  Emittance  Spectroscopy, Cambridge University press: 2005.

\bibitem[\protect\citeauthoryear{ Hapke  et al.}{1986}]{b1}  Hapke,B.W., Shepard ,M. K., Nelson, R. M., Smythe,W. D.,  Piatek ,J. L., 2009, Icarus, 199, 210-218

\bibitem[\protect\citeauthoryear{Henyey et al.}{1986}]{b1}  Henyey  C. and Greenstein J., 1941. ApJ, 93,70-83.

\bibitem[\protect\citeauthoryear{Kamei et al.}{1986}]{b1} Kamei A, 1999. Kogachi M., Mukai T.,Nakamura AM., Adv. Space Res, 23:1205-08.

\bibitem[\protect\citeauthoryear{Kaasalainen et al.}{1986}]{b1} Kaasalainen S., 2003, A\&A, 409, 765-69.

\bibitem[\protect\citeauthoryear{Lumme \& Bowell}{1981}]{b1} Lumme K. \& Bowell E., 1981, ApJ, 86, 11.

\bibitem[\protect\citeauthoryear{Mishchenko}{1981}]{b1} Mishchenko, M.I.,1994, J. Quant. Spectrosc. Radiat. Transfer 52, 95-110.

\bibitem[\protect\citeauthoryear{Mishchenko}{1981}]{b1} Mishchenko, M.I., Macke, A., 1997, J. Quant. Spectrosc.Radiat. Transfer, 57, 767-794.

\bibitem[\protect\citeauthoryear{Mishchenko}{1981}]{b1} Mishchenko,M.I., Dlugach, J. M., Yanovitskij, E. G., Zakharova, N. T.,1999, J. Quant. Spectrosc. Radiat. Transfer,  63, 409-432

 \bibitem[\protect\citeauthoryear{Nelson et al.}{1986}]{b1} Nelson R M.,Smythe WD., Spiker L. J., 2000. Icarus, 147, 545-58.

 \bibitem[\protect\citeauthoryear{Piatek et al.}{1986}]{b1} Piatek ,J. L., Hapke, B. W.,Nelson, R. M., Smythe,W. D., Hale,A.S.,2004, Icarus, 171, 531-545

\bibitem[\protect\citeauthoryear{Pollack et al.}{1986}]{b1} Pollack J., Cuzzi J., J. Atmos. Sci., 1980. 37, 868-81.

\bibitem[\protect\citeauthoryear{Shepard et al.}{1986}]{b1} Shepard M K., Helfenstein,P. 2007, JGR, 112, 17

\bibitem[\protect\citeauthoryear{Shukratov et al.}{1986}]{b1} Shukratov Y., Ovcharenko A., Zubco E., 2002, Icarus, 159, 396-416.

 \bibitem[\protect\citeauthoryear{van de Hulst et al.}{1986}]{b1}  van de Hulst H C., 1957. Light Scattering by Small Particles.,New York: Wiley; 1957.
\end{thebibliography}
\end{document}